\pdfoutput=1
\documentclass{svproc}

\usepackage[utf8]{inputenc}
\usepackage{subcaption}
\usepackage{float} 
\usepackage{xcolor} 
\usepackage{longtable}
\usepackage{amsmath}
\usepackage{amssymb}
\usepackage{graphicx}
\usepackage[sectionbib]{natbib}

\newcommand{\ul}[1]{\underline{#1}}
\newcommand{\uul}[1]{\underline{\underline{#1}}}
\newcommand{\alp}{^{(\alpha)}}

\newcommand{\Ma}{ {\uul{M}\alp}}

\newcommand{\Tk}{\mathrm{T}_s}

\usepackage{url}

\begin{document}
\mainmatter              
\title{Model-based Assessment of the Cruising Traffic\\\& Environmental Impact of Parking Restrictions}
\titlerunning{Environmental Impact of Parking Restrictions}  
%
\author{Alexandre NICOLAS, Nilankur DUTTA, and Léo BULCKAEN}
\authorrunning{Alexandre NICOLAS et al.} 
%
%
\institute{Institut Lumière Matière, Universite Claude Bernard Lyon 1 and CNRS,\\
6 rue Ada Byron,
69622 Villeurbanne Cedex, France.\\
\email{alexandre.nicolas@cnrs.fr}}

\maketitle              

\begin{abstract}
In many large metropolitan areas, cars cruising for parking significantly contribute to congestion and pollution. At the same time, parking restrictions are contemplated to encourage the use of greener transport alternatives. We give a short overview of a versatile modelling framework for parking search, which can be solved by both numerical simulations and theoretical developments, and we illustrate its applicability in Lyon, France. Then, we show how this framework can be leveraged to assess the environmental impact of parking restrictions, weighing the antagonistic effects of the thus-induced excess cruising vs. mode shift.
\keywords{urban traffic, parking search, graph theory,  pollution}
\end{abstract}

\section{Introduction}

In many large metropolitan areas, parking is a central issue for transport authorities as well as for individual drivers \citep{shoup2018parking}. The latter may spend several dozens of hours every year  searching for parking, according to INRIX survey data \citep{INRIX2017},
while  cars cruising for parking may represent more than 10\% of the total traffic 
in many large cities (e.g., 15\% in central Stuttgart \citep{hampshire2018share}) and thus significantly contribute to congestion and pollution in city centres.
In parallel, parking is increasingly regarded as a
lever to enforce transport policies.

In particular, over the last ten or twenty years, restrictions on the parking supply have been implemented, or are under active consideration, in various cities across Europe, including Amsterdam, Copenhagen, Lyon, Munich, Paris, Stockholm, and Zurich \citep{kodransky2010europe}. This begs the following question: are exhaust emissions increased by such (actual or hypothetical) restrictions because they make parking search longer on average, or reduced because the elasticity of the parking demand curbs trips by car?

We start by underlining the relevance of parking-related studies using the concrete example of curbside parking in Lyon, France. 
Then, after recapping the general modelling framework that we recently put forward for the parking search problem and the insight provided by Statistical Physics and Graph Theory, we show how these considerations can enable transport engineers to assess the impact of hypothetical parking-related measures.

\section{A first case study: On-street Parking in Lyon, France \label{sec:Lyon}}

\paragraph*{  Context.}
To set the problem in a concrete context, let us illustrate it with the example of Lyon, France, as of 2019. The city belongs to the second
urban area in France, with a municipal population around 500,000 people, and 
it had more than 80k on-street parking spaces, excluding garages and private parking. Curbside parking was either free or charged at one of two rates, depending on the zone; residential offers were available. Besides, the city has 
efficient transit services in the centre, with 4 metro lines, plus many tramway and bus lines.

\paragraph*{   On-street parking.}

\begin{figure}[!ht]
    \centering
   \includegraphics[width=0.36\linewidth]{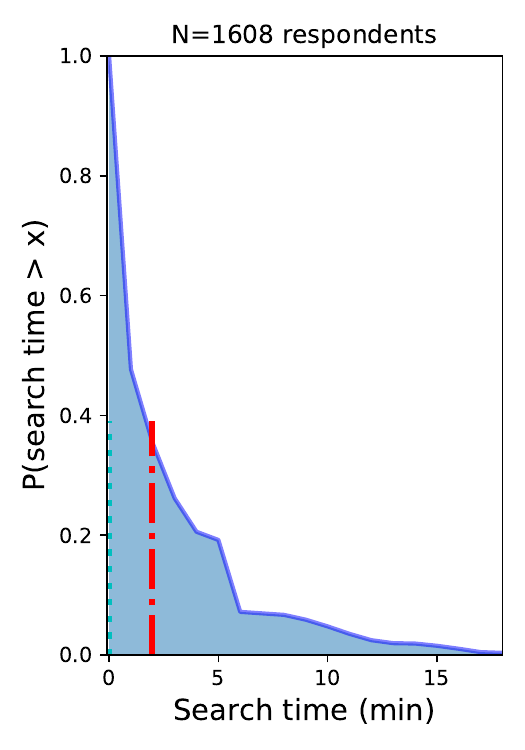}
   \includegraphics[width=0.6\linewidth]{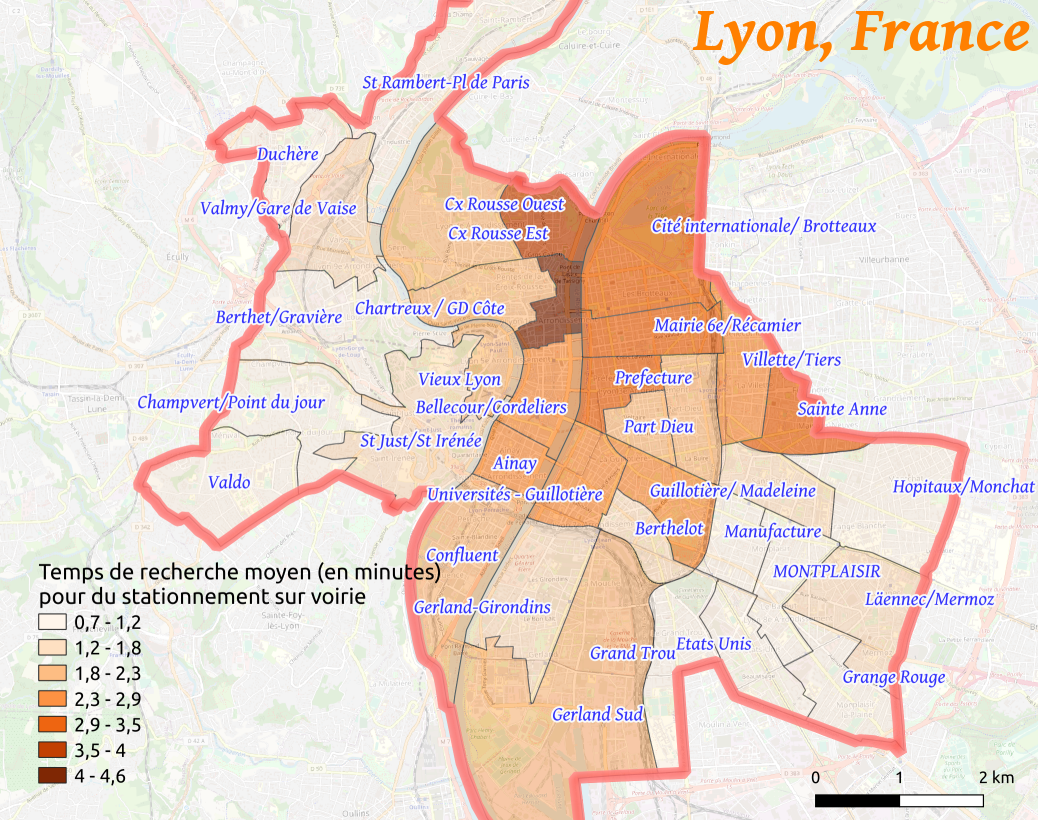}
        \caption{On-street parking search times in Lyon, France, as declared by respondents to the 2015 survey. (Left) Survival function of search times; vertical cyan and red lines represent the median and mean values, respectively. (Right) Average search time by geographic district. 
        Source of the data: \citep{Cerema2015EMD}.}
        \label{fig:search_times_Lyon_enquete}
\end{figure}

Difficulties to park in the city centre of Lyon have consistently been reported \citep{belloche2015street}. To evaluate them objectively, we  analyse the results of massive travel survey conducted by Cerema \citep{Cerema2015EMD} in 2015,
in which the respondents were asked about their parking search times, among other questions.
We focus on the 1,608 reports of on-street parking (regardless of the time of the day). The survival function $P(T\geqslant T_s)$ of parking search times $T_s$ is shown in Fig.~\ref{fig:search_times_Lyon_enquete}(left) and has a mean value between 2 and 3 minutes. While most motorists found a parking space almost instantly, a fraction of drivers had to cruise for more than 10 minutes, which drives upwards the mean search time. 
The issue of parking was considered a priority by many of the respondents.

\paragraph*{   Effect of parking restrictions.}

Given this context, one may wonder about the impact of various hypothetical modifications of the on-street parking plan, notably the (ongoing and prospective) restrictions on the parking supply. We purport to show that this question about hypothetical scenarios can be tackled using a generic methodology relying on a suitable modelling framework.

\section{Modelling Framework} 

We have recently introduced a modelling framework \citep{dutta2023parking} which encompasses several existing models \citep{benenson2008parkagent}. It
 takes as input a network
of streets, with parking spots located along them. Car drivers are
grouped into distinct categories $\alpha=1,\,2,\,\ldots$ depending on their destination, trip purpose, etc. At an intersection, they turn into an outgoing street 
with a probability given by a category-dependent turn-choice matrix 
${\uul{T}}\alp$, giving the probability that an $\alpha$-car moves along an edge of the graph in one arbitrary time step, \emph{if} it does not park in the meantime. When driving by a spot $i$, drivers
park there with probability $p_i\alp$ if it is vacant. Parked cars pull out at a rate $D\alp$, which is the reciprocal of the  average parking duration.

The on-site parking probability $p_i\alp$ may depend on a variety of explanatory variables (rate, distance to destination, etc.); here, these are subsumed into two generic variables: (i) an attractiveness $A_i\alp$ reflecting how attractive a spot $i$ is perceived to be \emph{intrinsically}, (ii) the driver's perception of how easy it currently is to park, $\beta\alp \in [0,\infty)$, viz., 
$p_i\alp(t)= f(A_i\alp,\beta\alp(t)) \in[0,1]$.
At very low occupancy, when parking seems extremely easy, the driver will refuse to park anywhere but in their preferred spot ($\beta\to \infty$). 
To the opposite, at extremely high occupancy, virtually any admissible spot 
will be deemed acceptable ($\beta \to 0$).

\paragraph*{  Theoretical insight.}

On top of its versatility, a major asset of the model is that it can be addressed not only by means of direct numerical simulations, but also more theoretically. 
For that purpose, the street network is handled as a graph, considering that every street position associated with a parking spot as well as every intersection are nodes of this graph.
The instantaneous number of cars of category $\alpha$, $\alpha$-cars,
passing by each node, averaged over random realisations, is then represented by a vector $\ul{V}\alp(t)$. The transition matrix $T_{ij}\alp$ is useful to locate cars
after an arbitrary number o steps, \emph{if they do not park in the mean-time}. However, to account for cars pulling in, $T_{ij}\alp$ is substituted by $M\alp_{ij}= (1-p_i\alp\,\hat{n}_i)\cdot T_{ij}\alp$, where  $\hat{n}_i=1-n_i=0$ if the spot is vacant (else, 1). 

The average `driving, searching, and parking' time $\Tk^{(\alpha,j)}$ of an $\alpha$-car  parking at spot $j$ (in arbitrary time steps) can then be derived \citep{dutta2023parking},
\begin{equation}
    \Tk^{(\alpha)}  = V_i\alp(0) \cdot  \Big[ \Ma\cdot 
                     \Big(\uul{\mathbb{I}} - \Ma\Big)^{-2}\Big]_{ij}\cdot  p_j\alp\,\hat{n}_j.
\label{eq:stime_alpha_spots}
\end{equation}

To get the mean occupancy $\ul{n}$, we resort to a mean-field approximation in the stationary regime; a conservation equation yields the desired mean-field stationary occupancy $\langle n_j \rangle$, as detailed in \cite{dutta2023parking}.

\paragraph*{  Application to Lyon.}
Next, we apply the foregoing generic framework to our example of on-street parking in Lyon in the morning peak hour (7am to 10am). Origin-Destination matrices were constructed based on population and employment data and parking occupancy data were measured locally; further details about the specification of the model are provided in \citep{dutta2023parking}. 

\begin{figure}[!htb]
    \centering
        \includegraphics[width=0.65\textwidth]{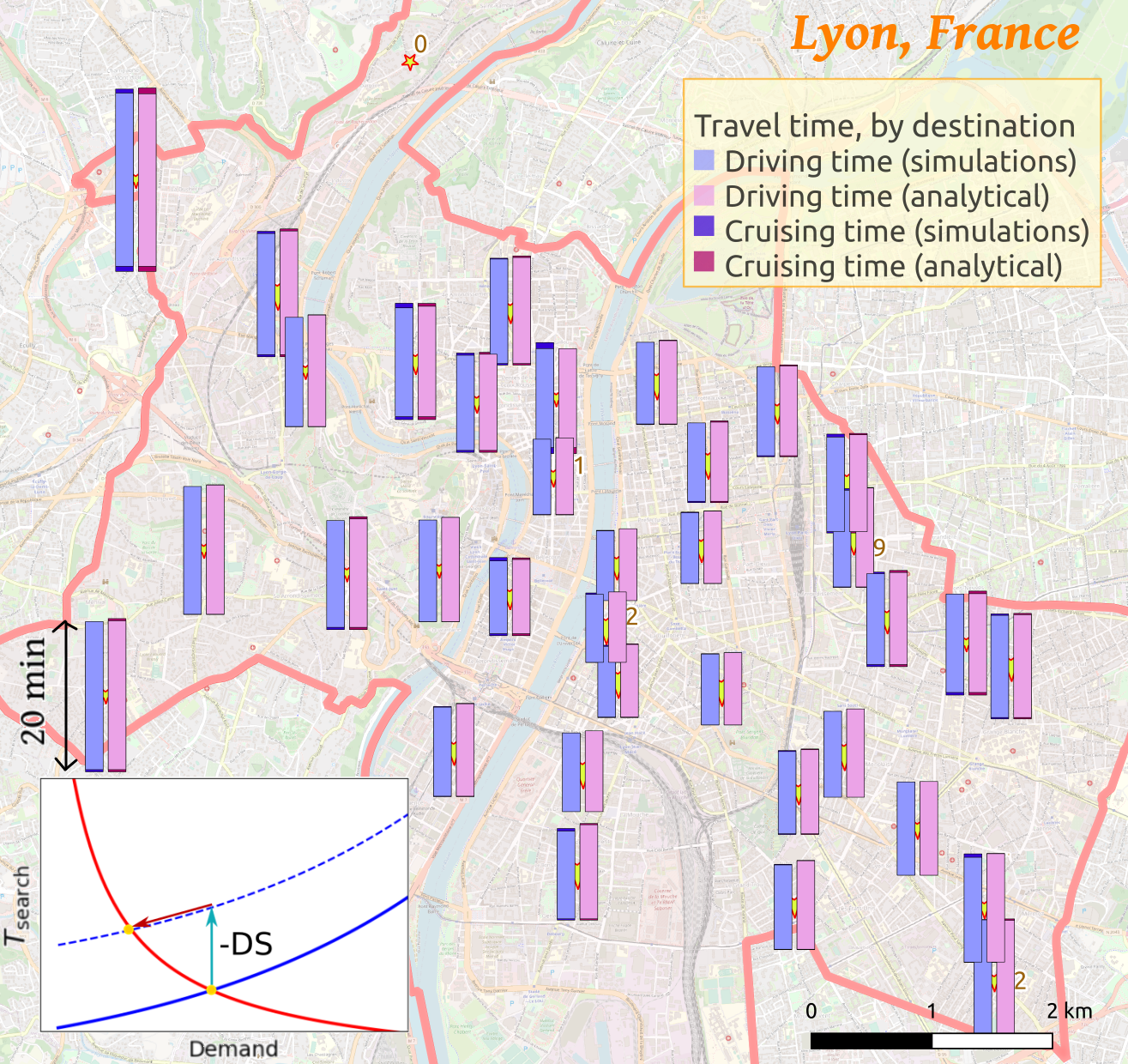}
\caption{Mean travel times (driving and cruising) in Lyon: comparison between simulations (in blue) and theory (in pink) for all  destinations, irrespective of the entry point, for a global injection rate of 30 cars/min. 
 (\emph{Inset}) Schematic  relation between the search time afforded by the street network and the parking demand, and the effect of a reduction $DS$ in parking supply. }
\label{fig:Lyon_stimes_dest}
\end{figure}

The  average total travel times (including parking search) simulated numerically are shown in Fig.~\ref{fig:Lyon_stimes_dest}, distinguished between destinations. Remarkably, no matter how complex the network is and how many different categories of drivers there are, the theoretical reasoning exposed above (Eq.~\ref{eq:stime_alpha_spots}) allows us to quantitatively reproduce most of the numerical results.
From the total travel time, the parking search time can be deduced by subtracting the travel time for \emph{vanishing} parking demand from the travel time with the \emph{actual} parking demand. We observe that these search times are also satisfactorily predicted by the analytical formulae  (Fig.~\ref{fig:Lyon_stimes_dest}), although some inaccuracies due to the mean-field approximation can be spotted. Still, globally, the zones of high parking tension are well captured by the model, except that it predicts more tension in the North West than what is actually reported.

\section{Environmental Impact of Parking Restrictions}

Armed with this framework, let us now address the question raised in the introduction,
namely, if reducing the parking supply (say, from $S$ parking spaces to $S'=S+\Delta S$, with $\Delta S<0$) will be environmentally noxious (in that
it generates additional cruising traffic) or beneficial (by curbing traffic). 
We will start by exposing 
the methodology in a deliberately very stylised way (somewhat akin to \citep{madsen2013model}), in which only \emph{one period} and one category of drivers (i.e., \emph{one destination} and trip purpose) are considered.

First, let us study the impact of this measure, \emph{ceteris paribus}, on the parking search time $T_{\mathrm{search}}(N,S)$, where the parking demand $N= N(T_{\mathrm{search}})$ is the number of cars aiming to park  around the destination:

\begin{equation}
    T_{s}(S')-T_{s}(S)
    \approx \Delta S\cdot\frac{dT_{s}}{dS}
    =\Delta S\cdot\left[\frac{\partial T_{s}}{\partial S}+\frac{\partial T_{s}}{\partial N}\cdot\frac{dN}{dS}\right] .
\end{equation}

Since $\frac{dN}{dS}= (\frac{N}{T_s} \epsilon_N) \cdot \frac{dT_{s}}{dS}$, where
$\epsilon_N:=\frac{\partial \ln N}{\partial \ln T_{s}}<0$ is the elasticity of parking demand to search time,
we arrive at

\begin{equation}
    \frac{dT_{s}}{dS} = \frac{1}{1+ \frac{N}{T_s} \cdot |\epsilon_N| \cdot \frac{\partial T_{s}}{\partial N}} \cdot \frac{\partial T_{s}}{\partial S}
\end{equation}
Here, we notice that the expected hike of $T_s$, $\partial T_s / \partial S$,
is mitigated by the dip in demand that it induces, via $\epsilon_N$.
If the public transit network is accessible and efficient, we expect a high elasticity $|\epsilon_N|$, hence a modest impact of the parking supply reduction on the search time because of mode shift.

Finally, the environmental impact of the measure is assessed by expressing the variation in the total exhaust emissions $V_p= V_p(N,T_s,\mathrm{traffic\ cond.})$ (neglecting the carbon footprint of alternative transport modes, for simplicity) as
\begin{eqnarray}
    \frac{dV_p}{dS} &=&  \frac{V_p}{N} \cdot \frac{dN}{dS} + \frac{\partial V_p}{\partial T_s}\cdot \frac{dT_s}{dS} + \mathrm{congestion\ effect} \\
    &=& \frac{dT_{s}}{dS} \cdot \Big[ 
    \underset{\mathrm{traffic\,evaporation}}{\underbrace{ \epsilon_N \frac{V_p}{T_s}} } +
    \underset{\mathrm{excess\,cruising}}{\underbrace{ \frac{\partial V_p}{\partial T_s}}} \Big] + \mathrm{congestion\ effect} 
\end{eqnarray}
Besides the possible (higher-order) backlash on congestion, there are two antagonistic effects between the square brackets.
The first one, the demand elasticity to search time, represents the
evaporation of part of the traffic following the measure, it indicates
how many motorists have switched to another transport mode or given up their journey altogether.
The second one is the adverse
effect due to the excess cruising generated by tighter parking supply. Both effects are magnified by the sensitiveness of
parking search times to these constraints.

\paragraph*{Caricatural examples}. Let us illustrate these findings with two antipodal,  caricatural examples: (A) Consider
a city with sprawling suburbs and extensive motorised commuting (large $V_p$) despite competitive transit services (large $|\epsilon_N|$), then the first term will largely prevail and the parking supply reduction will be environmentally beneficial, as often found empirically \citep{kodransky2010europe}.
(B) Conversely, should there be no competitive alternative to driving, the reduction in the parking supply is likely to increase pollutant emission, especially if the induced cruising may represent a significant portion of the total trip.

\paragraph*{Conclusion and broader applicability.}
In summary, thanks to a recently proposed quantitative model to estimate parking search times, we have been able to assess in generic terms the environmental impact of parking restrictions, depending on the motorists' propensity to modal shift; the results are qualitatively compatible with observations on travel practices in zones with different degrees of parking tension \citep{merle2009contrainte}.
Of course, practical cases will never be as clear-cut as the above two examples. More quantitative evaluations are therefore needed, and this is where the framework introduced above can be particularly instrumental, notably for the evaluation of the search time $T_s$ and its partial derivatives, even without resorting to intensive agent-based simulations. (In parallel, the elasticity $\epsilon_N$ can be estimated by local (e.g., phone) surveys.)
Moving towards these more realistic situations requires relaxing the one-driver-category assumption, but this is readily done in our model,
by replacing the scalars $N$, $T_s$, $S$ with vectors $\ul{N}$, $\ul{T_s}$, $\ul{S}$ whose components correspond to the destinations ($\alpha$). The idea can even be further generalised by  substituting the variation in parking supply $\ul{\Delta S}$ with modified
attractivenesses $A\alp_i$ for the individual parking spaces $i$, to account for changes in e.g. parking rates. These more realistic situations will be dealt with in an upcoming manuscript.

\paragraph*{Acknowledgement. }
We thank the City of Lyon for providing us parking data. This work was partly funded by an Impulsion grant of Université de Lyon. The maps on the figures were produced using QGIS (https://www.qgis.org).


\begin{thebibliography}{10}
\providecommand{\natexlab}[1]{#1}
\providecommand{\url}[1]{{#1}}
\providecommand{\urlprefix}{URL }
\expandafter\ifx\csname urlstyle\endcsname\relax
  \providecommand{\doi}[1]{DOI~\discretionary{}{}{}#1}\else
  \providecommand{\doi}{DOI~\discretionary{}{}{}\begingroup
  \urlstyle{rm}\Url}\fi
\providecommand{\eprint}[2][]{\url{#2}}

\bibitem[{Belloche(2015)}]{belloche2015street}
Belloche S (2015) On-street parking search time modelling and validation with
  survey-based data. Transportation Research Procedia 6:313--324

\bibitem[{Benenson et~al(2008)Benenson, Martens, and
  Birfir}]{benenson2008parkagent}
Benenson I, Martens K, Birfir S (2008) Parkagent: An agent-based model of
  parking in the city. Computers, Env and Urban Systems 32(6):431--439

\bibitem[{{CEREMA}(2015)}]{Cerema2015EMD}
{CEREMA} (2015) {Enqu\^ete m\'enages d\'eplacements (EMD), Lyon / Aire
  m\'etropolitaine lyonnaise}. ADISP (broadcaster)

\bibitem[{Cookson and Pishue(2017)}]{INRIX2017}
Cookson G, Pishue B (2017). Impact of parking pain in US, UK and Germany.

\bibitem[{Dutta et~al(2023)Dutta, Charlottin, and Nicolas}]{dutta2023parking}
Dutta N, Charlottin T, Nicolas A (2023) Parking search in the physical world
  [...]. Transportation Science 57(3):685--700

\bibitem[{Gragera et~al(2021)Gragera, Hybel, Madsen, and
  Mulalic}]{madsen2013model}
Gragera A, Hybel J, Madsen E, Mulalic I (2021) A model for estimation of the
  demand for on-street parking. Economics of Transportation 28:100,231

\bibitem[{Hampshire and Shoup(2018)}]{hampshire2018share}
Hampshire RC, Shoup D (2018) What share of traffic is cruising for parking?
  Journal of Transport Economics and Policy (JTEP) 52(3):184--201

\bibitem[{Kodransky and Hermann(2010)}]{kodransky2010europe}
Kodransky M, Hermann G (2010) Europe's parking U-Turn: From accomodation to
  regulation. Institute for Transp. and Dev. Policy New York, USA

\bibitem[{Merle and Verry(2009)}]{merle2009contrainte}
Merle N, Verry D (2009) Contrainte de stationnement et pratiques modales:
  m{\'e}thodologie et {\'e}tude des cas de lille, lyon et montpellier. Tech.
  rep., Certu

\bibitem[{Shoup(2018)}]{shoup2018parking}
Shoup D (2018) Parking and the City. Routledge

\end{thebibliography}
\end{document}